
\magnification=1200
\hyphenpenalty=2000
\tolerance=10000
\hsize 14.5truecm
\hoffset 1.truecm
\openup 5pt
\baselineskip=24truept
\font\titl=cmbx12
\def\der#1#2{{\partial#1\over\partial#2}}


\def\a{^\prime}
\def\rg{r_g}

\def\ref{\par\noindent\hangindent 20pt}

\def\mincir{\raise -2.truept\hbox{\rlap{\hbox{$\sim$}}\raise5.truept
\hbox{$<$}\ }}
\def\magcir{\raise -4.truept\hbox{\rlap{\hbox{$\sim$}}\raise5.truept
\hbox{$>$}\ }}
\def\magmin{\raise -4.truept\hbox{\rlap{\hbox{$<$}}\raise5.truept
\hbox{$>$}\ }}
\def\minmag{\raise -4.truept\hbox{\rlap{\hbox{$>$}}\raise5.truept
\hbox{$<$}\ }}
\def\rho{\varrho}

\null\vskip 1.truecm\parindent 0pt
\centerline{\titl{ON THE MATHEMATICAL CHARACTER OF THE}}
\medskip
\centerline{\titl{RELATIVISTIC TRANSFER MOMENT EQUATIONS}}
\vskip 1.5truecm
\vskip 1.5truecm
\centerline{R.~Turolla $^1$, L.~Zampieri $^2$ and L.~Nobili $^1$}
\bigskip\bigskip
\centerline{$^1$ Department of Physics, University of Padova}
\smallskip
\centerline{Via Marzolo 8, 35131 Padova, Italy}
\bigskip
\centerline{$^2$ International School for Advanced Studies}
\smallskip
\centerline{Via Beirut 2--4, 34014 Trieste, Italy}
\vfill
\centerline{Accepted for publication in MNRAS}
\bigskip
\eject
\parindent 1.truecm

\beginsection ABSTRACT

General--relativistic, frequency--dependent radiative transfer in spherical,
differentially--moving media is considered. In particular we investigate
the character of the differential operator defined by the first two moment
equations in the stationary case. We prove that the moment equations
form a hyperbolic system when the logarithmic velocity gradient is positive,
provided that a reasonable condition on the Eddington factors is met.
The operator, however, may become elliptic in accretion flows and, in
general, when gravity is taken into account. Finally we show that,
in an optically thick medium, one of the characteristics becomes infinite
when the flow velocity equals $\pm c/\sqrt 3$. Both high--speed, stationary
inflows and outflows may therefore contain regions which are ``causally''
disconnected.
\bigskip\noindent
Key Words: Radiative Transfer \ -- \ Relativity

\beginsection 1. INTRODUCTION

Radiative transfer in differentially--moving media has been extensively
investigated in the past and a large body of literature is available on
this subject (see Mihalas \& Mihalas 1984, and references therein). Despite
the large efforts, however, works dealing with the transfer of radiation
through media moving at relativistic speeds are comparatively few. The special
relativistic transfer equation was firstly derived by Thomas (1930) and,
including Thomson scattering, by Simon (1963) and Castor (1972); a thorough
derivation can be found in the monograph by Mihalas \& Mihalas.
Stationary solutions in spherical symmetry were discussed by
Mihalas (1980), Mihalas, Winkler \& Norman (1984) and Hauschildt \& Wehrse
(1991). Radiative transfer in curved spacetimes was investigated
by Lindquist (1966), Anderson \& Spiegel (1972), Thorne (1981), Schinder
(1988),
Schinder \& Bludman (1989), Anile \& Romano (1992) and Nobili, Turolla \&
Zampieri (1993). All the solutions to the relativistic transfer problem
found up to now were obtained using essentially two different approaches:
either the transfer equation is directly solved for the radiation
intensity $I_\nu(r,\mu )$ or the angular
dependence is removed by introducing the moments of $I_\nu(r,\mu )$ and
the moment equations are then integrated. Each method has both advantages
and disadvantages. The solution of the transfer equation gives directly
$I_\nu(r,\mu )$ but is a quite
formidable numerical task and requires special techniques, like the
DOME method introduced by Hauschildt \& Wehrse. On the other hand,
the moment equations, being of lower dimensionality, are easier to handle
numerically but their solution alone does not specify completely
$I_\nu(r,\mu )$. Moreover, an exact solution for the moments themselves
can be obtained only if the Eddington factors $f_E = K_\nu/J_\nu -1/3$ and
$g_E = N_\nu/H_\nu -3/5$ are computed self--consistently
(these are the definitions of $f_E$ and $g_E$ appropriate to the PSTF
moment formalism, see below). An exact determination of
$f_E$ and $g_E$ was
obtained by means of the Tangent Ray Method (TRM, originally due to
Mihalas, Kunasz \& Hummer 1975) by Mihalas (1980) in the special relativistic
case and by Schinder \& Bludman (1989) for a spherically--symmetric, static
atmosphere in a Schwarzschild spacetime. While it is not entirely obvious
that TRM could be fruitfully applied to more complex situations in which
dynamics and gravity are both present, the moment method can still be used
to get,
at least, an approximate solution for $I_\nu(r,\mu )$ by introducing ,``a
priori'', reasonable expressions for the Eddington factors (see e.g. Minerbo
1978, Nobili {\it et al.\/} 1993).

In this paper we present an analysis of the stationary, spherically--symmetric
relativistic moment equations placing
particular emphasis on the character of the second order differential
operator implicitly defined by the zero--th and first equations. This point
never received proper attention in the past, despite the fact that it
appears to be of non--negligible relevance since the choice of the
boundary conditions is crucially related to the character of the operator.
Mihalas, Kusnaz \& Hummer (1976) discussed to some extent the problem of
frequency conditions in the non--relativistic case, concluding that for
accelerated winds and decelerated inflows the transfer equation is of the
Feautrier type, although
they stressed that different velocity laws may produce anomalous behaviours.
Here we prove that the special relativistic moment equations form a hyperbolic
system for positive logarithmic velocity gradients at least if $f_E -
g_E -4/15<0$,
so that they should be solved as a two--point boundary value problem in space
and an initial value problem in frequency. In converging flows, however,
advection and aberration effects may produce a region of ellipticity.
Moreover, when gravity is taken into account, the operator may become elliptic
even in the wind case. We point out that one of the two characteristics,
when they exist, diverges when the flow velocity equals a ``sound'' speed
$v_s$, $1/3\leq (v_s/c)^2\leq 1$ depending on optical depth. The sphere of
radius $r_s$ is completely analogous to a sound horizon and behaves like
a one--way membrane as far as the propagation of boundary data is concerned.
Finally the connection between the ``pathologies'' in the moment equations
and the vanishing of the coefficients of $I_\nu$--derivatives in the transfer
equation is discussed.

\beginsection 2. MOMENT EQUATIONS

General relativistic moment equations were derived by Thorne (1981) who
introduced Projected Symmetric Trace Free tensors to describe the moments
of the radiation intensity. In spherical symmetry the $k$--th PSTF moment has
just one independent component, the radial one denoted by $w_k$, and the
formalism greatly simplifies. The first two radiation moments are the
radiation energy density and flux
measured by a comoving observer; the third PSTF moment is the radiative
shear, $w_2 = 4\pi (K_\nu - J_\nu/3)$. An alternative form of the moment
equations in a static, spherical spacetime, making use of a Lagrangian comoving
co--ordinate system, has been presented by Schinder and Schinder \& Bludman.

Using Thorne's notation, with all moments in ergs cm$^{-3}$, the first
two stationary moment equations are (see also Nobili {\it et al.\/})
$$
\eqalignno{&
\der{w_1}{\ln r} +   2w_1 + {{y\a}\over{y}} \left(
w_1 - \der{w_1}{\ln\nu} \right)
 + {v\over c} \left[ - \left( {{u\a}\over{u}} - 1\right) \der{w_2}{\ln\nu}
\right. & \cr
& \left. + \der{w_0}{\ln r}
 + \left(2 + {{u\a}\over{u}}\right)w_0
-{1\over 3} \left( 2 + {{u\a}\over u}\right) \der{w_0}{\ln\nu} \right]
= {{s_\nu^0 r}\over{y}}
& (1a)\cr }
$$

$$\eqalignno {&
{1\over 3}\der{w_0}{\ln r}
+ {{y\a}\over{y}}\left(w_0 - {1\over 3}\der{w_0}{\ln\nu}\right)
+ \der{w_2}{\ln r}
+ 3w_2 - {{y\a}\over{y}}\der{w_2}{\ln\nu} &\cr
& + {v\over c}\left[\der{w_1}{\ln r}
 + {1\over 5}\left(7{{u\a}\over{u}} + 8 \right)w_1
 -{1\over 5}\left(3{{u\a}\over{u}} +2\right)
\der{w_1}{\ln\nu} -
  \right. &\cr
& \left. \left({{u\a}\over{u}}
-1\right)\left( w_3+ \der{w_3}{\ln\nu}\right)\right]
={{s_\nu^1 r}\over{y}}.&(1b)\cr}$$
Here $y = \gamma\sqrt{1-\rg/r}$ is the total energy per unit mass, $u=yv/c$ is
the radial component of the fluid 4--velocity and $s_\nu^0$, $s_\nu^1$ are the
source moments; a prime denotes the total derivative in the $r$--direction and
$r_g$ is the gravitational radius.

An analysis on the nature of the various dynamical terms appearing in the
moment equations was presented by Buchler (1983); a similar discussion for the
transfer equation can be found in Castor (1972), Mihalas {\it et al.\/}
(1976), Hauschildt \& Wehrse (1991). Terms of order $v/c$ in equations (1)
account both for the local Doppler shift of photons and for advection and
aberration. Mihalas {\it et al.\/} have shown that advection and
aberration produce a
fractional variation on the solution which is $\sim 5 v/c$ and
can be safely neglected for small velocities.
In some astrophysical situations however, like photospheric supernovae
expansion, jets, accretion onto black holes and neutron stars,
velocities $\magcir 0.1 c$ are expected and such effects cannot be ignored.
In the next sections we show that, apart from obvious quantitative effects,
advection/aberration terms may change substantially the mathematical
properties of the moment equations in spherical inflows.

\beginsection 3. CHARACTERISTIC ANALYSIS

In the following we investigate the mathematical character of the second
order differential operator defined implicitly by equations (1). For the
sake of simplicity we shall assume that source moments
contain no derivatives of the radiation moments; an extension of the present
analysis will be needed to include radiative processes like
non--conservative electron scattering treated in the Fokker--Planck
approximation which depends on both first and second $\nu$--derivatives of
$w_0$ (see discussion at the end of this section).

The characteristic analysis of a generic, linear system of first order partial
differential equations can be easily performed once the system is brought
into the form (see e.g. Whitham 1974)
$$\der{u_i}{t} + a_{ij}\der{u_j}{x} + b_i(x,t;u_j) = 0\qquad\qquad i,j=1,n.
\eqno(2)$$
In this case the characteristic velocities are the roots of the equation
$$\vert a_{ij} - \lambda\delta_{ij}\vert =0\eqno(3)$$
and the system is hyperbolic if equation (3) has $n$ different real roots.
Rewriting the moment equations in the form (2) and introducing the Eddington
factors $f_E = w_2/w_0 = K_\nu/J_\nu -1/3$, $g_E=w_3/w_1=N_\nu/H_\nu -3/5$,
we obtain, after some manipulations,
\smallskip
$$\eqalignno{ &
\der{w_0}{\ln r} + {{1}\over{f_E+1/3-v^2/c^2}}\left\{\left[
{{v^2}\over {c^2}}F-{{y\a}\over {y}}\left(f_E+{1\over 3}-{{v^2}\over {c^2}}
\right)\right]
\der{w_0}{\ln\nu} \right. &\cr
&\left. - {v\over c} G\der{w_1}{\ln\nu}\right\} + C_1 = 0 &(4a)\cr}$$

$$\eqalignno{ &
\der{w_1}{\ln r} + {1\over{f_E+1/3-v^2/c^2}}\left\{ - {v\over c}
\left(f_E + {1\over 3}\right)F
\der{w_0}{\ln\nu}+ \right. &\cr
&\left.
\left[{{v^2}\over {c^2}}G-{{y\a}\over {y}}\left(f_E+{1\over 3}-{{v^2}\over
{c^2}}\right)
\right]\der{w_1}{\ln\nu}\right\} + C_2 = 0. &(4b)\cr}$$
Here $F = (\beta -1)f_E + (2+\beta )/3 - y\a/y$, $G = (\beta -1)g_E +
(2+3\beta )/5 - y\a/y$,
$\beta = u\a /u$, $y\a/y = \beta v^2/c^2 + r_g/2y^2r$
and all terms not containing
derivatives of the moments are grouped together into $C_1$ and $C_2$.
Actually $C_1$ and $C_2$ do contain derivatives of both $f_E$ and $g_E$
but the Eddington factors are to be regarded as known functions either coming
from the solution of the transfer equation, as in the TRM,
or being specified ``a priori'' if a closure is assumed.
The Eddington factors used here differ from the usual ones inasmuch PSTF
moments are originated by a Legendre polynomial expansion of the intensity;
in particular $f_E$ and $g_E$ are restricted in the ranges $0\leq f_E\leq 2/3$
and $0\leq g_E\leq 2/5$.
We note that the matrix of the coefficients $a_{ij}$, defined in equation (2),
is symmetric and then its eigenvalues are real,
if $f_E = 2/3$ and $g_E = 2/5$. This means that the moment equations are
always hyperbolic in the streaming limit, for any value of $v$ and $\beta$.

The equation for the characteristic velocities is
$$\eqalignno{ &
\lambda^2 + {1\over{f_E+1/3-v^2/c^2}}\left[
2{{y\a}\over {y}}\left(f_E+ {1\over 3} - {{v^2}\over{c^2}} \right)
- {{v^2}\over{c^2}}(F + G) \right]\lambda &\cr
&
+ {1\over{f_E+1/3-v^2/c^2}}\left[\left({{y\a}\over {y}}\right)^2
\left(f_E + {1\over 3} - {{v^2}\over{c^2}} \right) \right. &\cr
& \left. - {{y\a}\over {y}} {{v^2}\over{c^2}} \left( F+G \right)
- {{v^2}\over{c^2}}FG\right]=0. &(5)\cr}$$
By introducing $U^2 = (v^2/c^2)/(f_E + 1/3)$, the discriminant of equation
(5) can be written as
$$\eqalignno{\Delta & = {{U^2}\over{(U^2 - 1)^2}}\left[(U^2 - 1)(F-G)^2+
\left(F+G\right)^2\right ] &\cr
&={{U^2}\over{(U^2 - 1)^2}}\left[U^2 (F-G)^2+ 4FG\right ]\, .
& (6)\cr}$$

{}From equation (6) it follows that the sign of $\Delta$ depends only on the
sign of the term in square brackets and it is
$\Delta>0$ if either $U^2 >1$, that is to say $v^2/c^2>f_E+1/3$, or $FG>0$,
regardless of the values of the Eddington
factors and of the velocity gradient. In order to make the analytical treatment
affordable in the following we shall neglect gravity, so that $y\a/y =
\beta v^2/c^2$.
In this case it is easy to see that $F$ and $G$ are opposite in sign
and, consequently, $\Delta$ may become negative for flow velocities in the
range $a < v^2/c^2 < b$, where

$$ \eqalign{
& a = \min \left[ f_E + {1\over 3} + {1\over\beta}\left({2\over 3} - f_E\right
)
\, , \,
g_E + {3\over 5} + {1\over\beta}\left({2\over 5} - g_E\right)\right] \cr
& b = \max \left[ f_E + {1\over 3} + {1\over\beta}\left({2\over 3} - f_E\right)
\, , \,
g_E + {3\over 5} + {1\over\beta}\left({2\over 5} - g_E\right) \right] \, .\cr}
 (7)  $$
Let us consider the case $\beta >0$ first. Since we have already shown that
$\Delta>0$ if $v^2/c^2>f_E+1/3$, it follows that only the velocity interval

$$ g_E + {3\over 5} + {1\over\beta}\left({2\over 5} - g_E\right) <
{{v^2}\over{c^2}} <
f_E + {1\over 3} \eqno(8) $$
needs to be considered. It can be easily checked that the above conditions are
never fulfilled if $0<\beta \leq 1$. For $\beta > 1$, a sufficient condition
for the positiveness of $\Delta$ can be obtained imposing that

$$ g_E + {3\over 5} + {1\over\beta}\left({2\over 5} - g_E\right) >
f_E + {1\over 3}\, , $$
which is equivalent to

$$(\beta -1)(g_E-f_E+ {4\over{15}}) + {2\over 3} - f_E>0\, .$$
In order for the left hand side of the previous inequality to be positive,
it is enough to ask that
$$f_E - g_E -{4\over {15}}<0 \eqno(9)$$
which, we stress again, gives only a {\it sufficient\/} conditions for the
hyperbolicity of the moment equations for $\beta >1$. On the other hand, we
note that if condition (9) is violated there always exists a value of
$\beta$, $\beta = 1 + (2/3 - f_E)/(f_E - g_E - 4/15)> 1$, beyond which $\Delta$
may become negative.

Condition (9) can not be proved to hold in full generality and should be
verified case by case. It should be taken into account, however, that the first
two Eddington factors are not independent from each other, although we avoided
up to now to specify any relation between them. In order to check if condition
(9) can be physically acceptable, we compare it with the results obtained by Fu
(1987a, b). Using a statistical formalism to approximate the radiation
intensity at all depths, he was able to derive constraints on the Eddington
factors, showing that the values of $K/J$ are bounded by two curves, the
``logarithmic'' (upper) and ``hyperbolic'' (lower) limits, in the $H/J\times
K/J$ plane. Expressed in terms of the more conventional Eddington factors $K/J$
and $N/H$, condition (9) reduces to $K/J < N/H$. We have computed the
logarithmic and hyperbolic limits of the second Eddington factor and
verified that it is $(K/J)_{log}\leq (N/H)_{log}$, $(K/J)_{hyp}\leq
(N/H)_{hyp}$ for $H/J\leq 1$, although the inequality $(K/J)_{log}\leq (
N/H)_{hyp}$ is not satisfied for $H/J >0.67$. On the other hand, since the
hyperbolic (logarithmic) limit should be attained in the streaming
(diffusion) regime, it seems more meaningful to compare values of $K/J$
and $N/H$ that describe statistical properties of the radiation field in
the same physical conditions; so our request that $K/J < N/H$ at all depths
seems indeed compatible with the results of Fu's analysis.

Unfortunately the study of the limits given by equation (7) is not so
straightforward if $\beta<0$ and when gravity is taken into account.
However, if the gravitational field is strong enough and/or the gas flow
is almost in free--fall,
the existence of velocity ranges where $\Delta$ changes sign, and
the operator becomes elliptic,
is certainly possible, even if $v/c$ is small.

Up to now we have discussed the conditions for the existence of real
characteristics without considering the actual behaviour of the characteristics
themselves. Assuming $\beta >0$, $f_E - g_E - 4/15 <0$ and neglecting gravity,
equation (5) can be
used to analyze how the characteristics change varying $v/c$. In the
limit of vanishing velocity there is just the double root $\lambda
=0$ which indicates that the two moment equations decouple (no ``frequency
mixing'' between the moments). As $v/c$ increases the characteristics become
distinct. It is possible to prove that the solutions of equation (5) are
opposite in sign, but not equal in magnitude, if $(v/c)^2< f_E + 1/3$. From
equation (5), in fact, it follows that the product of the roots, $\lambda_1
\lambda_2$, is
$$\eqalign{\lambda_1 \lambda_2 = & {{v^2/c^2}\over{f_E+1/3-v^2/c^2}}\left[
\beta^2\left(f_E + {1\over 3} - {{v^2}\over {c^2}} \right){{v^2}\over{c^2}}
- \right.\cr
& \left.\beta \left( F+G \right) {{v^2}\over {c^2}}
- FG\right]\, .\cr}\eqno(10)$$
The term in square brackets can be written as

$$ \eqalign{& \beta\left[ \beta \left( f_E + {1\over 3} - {{v^2}\over {c^2}}
\right) - F \right] {{v^2}\over {c^2}}
- G \left( \beta {{v^2}\over {c^2}} + F \right) = \cr
& -\left(\beta {{v^2}\over {c^2}} + G\right) \left({2\over 3} - f_E\right) -
\beta \left( f_E + {1\over 3} \right) G\, . \cr } $$
Since $G>0$ for $(v/c)^2\leq f_E + 1/3$ (see condition [9]), the previous
expression is always negative and we can conclude that $\lambda_1 \lambda_2
< 0$ for $(v/c)^2 < f_E + 1/3$.

As equation (5) shows, one of the characteristics switches from
positive to negative through a pole at $(v/c)^2= f_E + 1/3$.
The existence of a diverging characteristic implies that the two spatial
regions separated by the line $r=r_s$, where $(v/c)^2= f_E + 1/3$, are
causally disconnected in the sense that the behaviour of the solution for
$(v/c)^2< f_E + 1/3$
is not influenced by what happens for $(v/c)^2> f_E + 1/3$. The surface of
radius $r_s$ behaves like a one--way membrane in the same way as the sound
horizon does in transsonic flows. As a consequence, if the flow
velocity equals the ``sound'' speed $v_s = (f_E + 1/3)^{1/2}c$, the moment
equations are not to be solved as a two--point boundary value problem in
space and an initial value problem in frequency, contrary to the case in
which $v<v_s$ everywhere.
Now the solution depends only on the data assigned on the spatial
boundary of the ``subsonic'' region plus the two initial frequency conditions.

The presence of a ``sound'' horizon is not an artifact introduced by the moment
expansion as can be seen examining the special relativistic form of the
transfer
equation, see e.g. Hauschildt \& Wehrse equation (1). The $r$--derivative
of the radiation intensity is, in fact, multiplied by the factor $e=\gamma
(\mu + v/c)$ which is  zero at $v/c=-\mu$. This means that if a
non--vanishing velocity field is present, the transfer equation becomes
necessary singular on the surface $v(r)/c = -\mu$ in the $r\times\mu\times\nu$
space, where the dimensionality of the equation lowers from 3 to 2. Moment
equations contain
the same kind of pathology but, being obtained by angle averaging the transfer
equation, the coefficients of the space derivatives vanish at a fixed value of
$\mu$ which is just $\sqrt{\langle\mu^2\rangle} = 1/\sqrt 3$ in the Eddington
approximation. We note that the transfer
equation does not exhibit any singularity when written in its characteristic
form in the $r\times\mu$ plane, as in the Tangent Ray Method, because this
amounts to use a coordinate system which establishes a one--to--one,
regular map between the integral surface and the integration domain.

The same kind of considerations can be used to relate the possible ellipticity
of the moment equations to the vanishing of the
coefficient of the $\nu$--derivative in the transfer equation,

$$g = {{\gamma}\over r} {v\over c}[1 - \mu^2 + \mu(\mu + v/c)\beta]\, ,
\eqno(11)$$
see equation (3c) of Hauschildt \& Wehrse where our definition of $\beta$
was used. It can be shown that $g$ is always non-negative for $-1\leq\mu\leq 1$
only if $ v > 0$ and $0\leq\beta\leq 2$; for all other values of the velocity
and of the velocity gradient there exist a value of $\mu$ at which $g$
vanishes. The transfer equation may, therefore, become singular even in the
outflow case and this agrees with the fact that the moment equations can be
proved to be hyperbolic without any additional constraint only for
$0\leq\beta\leq 1$. In general the degeneracy occurs along certain lines in
the $r\times\mu$ plane.
Actually, in the moment equations (4), the coefficients of the
$\nu$--derivatives, that are obviously related to $g$, contain
some averaged value of $\mu$ and they can change sign at a certain value
of $r$ in the integration domain. In particular, since these coefficients
depend on the two Eddington factors $f_E$ and $g_E$, they can change sign
at two different radii, say $r_1$ and $r_2$. This kind of
pathology manifests through the appearance of an interval ($r_1\, ,r_2$),
in which the system of differential equations becomes elliptic.

As stressed by Mihalas {\it et al.\/} (1976) and Hauschildt
\& Wehrse,  both $e$ and $g$ depend on the flow velocity only if advection
and aberration are taken into account, even to first order in $v/c$. In this
respect it is interesting to note that the moment equations reduce to a
parabolic system in the diffusion limit for any given $\beta$ if only local
Doppler shift of photons is retained (Nobili {\it et al.\/}, see also
Blandford \& Payne 1981 a, b, Payne \& Blandford 1981). It is, therefore,
the inclusion of advection and aberration terms, which act as singular
perturbations, that introduces
pathologies either in the transfer equation or in the system of the moment
equations. A similar conclusion, although in a different context, was reached
recently by Gombosi {\it et al.\/}
(1993) who studied energetic particle transport by means of a moment expansion
of the distribution function which is very similar to the one used here for
the radiation intensity.
A situation like this arises also when non--conservative
scattering is included in the source term. Assuming that it can be treated in
the Fokker--Planck approximation, the presence of $\nu$--derivatives of the
radiation intensity produces an effect analogous to advection/aberration. In
this case, see Colpi (1988), it can be shown that the transfer equation, in
the diffusion limit and retaining only local Doppler shift, is always of the
elliptic type and it must be integrated giving suitable conditions
on all the boundary of the integration domain.
Actually, we want to stress that a general
analysis of the mathematical character of the transfer equation is not
possible ``a priori'', depending on the input physics included in the source
term. In the present study, we dealt with the more complete form
of the moment equations in dynamical flows, but assuming that only
conservative scattering and isotropic true emission--absorption processes are
present.
Even under these assumptions, we have shown that in accretion flows the
operator defined by equations (1) may become of the mixed type, switching
from hyperbolic to elliptic. The presence of a spatially--limited elliptic
region around $\tau\approx 1$ implies that, there, conditions
must be specified at both the frequency boundaries, although
the problem remains two--point boundary valued in space.

As a final point, let us briefly discuss the effects induced by the
presence of a gravitational field on the existence of real characteristic
and, consequently, on the character of the operator defined by the moment
equations. Both the expressions for $F$ and $G$ contain, now, an extra term
$ - r_g/2 y^2 r$ with respect to the special relativistic case
and, even if $0\leq\beta\leq 1$, it could
be $\Delta > 0$ or $\Delta  < 0$, depending on the sign of $F =  F_{SR} - r_g/
2 y^2 r$ and $G= G_{SR} - r_g/2 y^2 r$.
This leads to the conclusion that, irrespective of the sign of $\beta$,
the presence of a gravitational field can induce a change in the character
of the moment equations; in particular, if $0\leq\beta\leq 1$
the possible appearance of regions of ellipticity is completely due
to gravity.

\beginsection 4. CONCLUSIONS

We have analyzed the mathematical character of the system formed by the first
two relativistic transfer moment equations. It has been shown that, similarly
to the non--relativistic case, the differential operator is of the hyperbolic
type when the flow velocity is a monotonically increasing function of the
radial
coordinate. On the contrary, in converging flows and when gravity is taken
into account, the character of the operator is
much more complex and the system of equations may become of the mixed type.
This result can be of interest in connection with models of spherical accretion
onto compact objects and seems to be originated by advection and aberration
effects.

\vfill\eject

\beginsection References

\ref{Anderson, J.L. \& Spiegel, E.A. 1972, ApJ, 171, 127}
\ref{Anile, A.M., \& Romano, V. 1992, ApJ, 386, 325}
\ref{Blandford, R.D., \& Payne, D.G. 1981a, MNRAS, 194, 1033}
\ref{Blandford, R.D., \& Payne, D.G. 1981b, MNRAS, 194, 1041}
\ref{Buchler, J.R. 1983, J. Quantit. Spectros. Radiat.
Transfer, 30, 395}
\ref{Castor, J.I. 1972, ApJ, 178, 779}
\ref{Colpi, M. 1988, ApJ, 326, 223}
\ref{Fu, A. 1987a, ApJ, 323, 211}
\ref{Fu, A. 1987b, ApJ, 323, 227}
\ref{Gombosi, T.I., Jokipii, J.R., Kota, J., Lorencz, K., \& Williams,
L.L. 1993, ApJ, 404, 377}
\ref{Hauschildt, P.H. \& Wehrse, R. 1991, J. Quantit. Spectros. Radiat.
Transfer, 46, 81}
\ref{Lindquist, R.W. 1966, Ann. Phys. (NY), 37, 487}
\ref{Mihalas, D., Kusnaz, P.B., \& Hummer, D.G. 1975, ApJ, 202, 465}
\ref{Mihalas, D., Kusnaz, P.B., \& Hummer, D.G. 1976, ApJ, 206, 515}
\ref{Mihalas, D. 1980, ApJ, 237, 574}
\ref{Mihalas, D., Winkler, K--H., \& Norman, M.L. 1984, J. Quantit. Spectros.
Radiat. Transfer, 31, 479}
\ref{Mihalas, D. \& Mihalas, B. 1984, Foundations of Radiation Hydrodynamics
(Oxford: Oxford University Press)}
\ref{Minerbo, G.N. 1978, J. Quantit. Spectros. Radiat. Transfer, 20, 541}
\ref{Nobili, L., Turolla, R., \& Zampieri, L. 1993, ApJ, 404, 686}
\ref{Payne, D.G., \& Blandford, R.D. 1981, MNRAS, 196, 781}
\ref{Schinder, P.J. 1988, Phys. Rev. D, 38, 1673}
\ref{Schinder, P.J. \& Bludman, S.A. 1989, ApJ, 346, 350}
\ref{Simon, R. 1963, J. Quantit. Spectros. Radiat. Transfer, 3, 1}
\ref{Thorne, K.S. 1981, MNRAS, 194, 439}
\ref{Whitham, G.B. 1974, Linear and Nonlinear Waves (New York: John Wiley
\& Sons)}

\vfill\eject\bye